\documentstyle[epsfig,12pt]{article}  
\newcommand{\beq}{\begin{equation}}
\newcommand{\eeq}{\end{equation}}
\newcommand{\bea}{\vspace{0.25cm}\begin{eqnarray}}
\newcommand{\eea}{\end{eqnarray}}

\newcommand{\rb}{\mbox{{\bf
r}}}

\def\lsim{\mathrel{\rlap{\lower4pt\hbox{\hskip1pt$\sim$}}
    \raise1pt\hbox{$<$}}}         %less than or approx. symbol
\def\gsim{\mathrel{\rlap{\lower4pt\hbox{\hskip1pt$\sim$}}
    \raise1pt\hbox{$>$}}}         %greater than or approx. symbol
\begin{document}
\vspace*{-2cm}
 
\bigskip
%%%%%%%%%%%%%%%%%%%%%%%%%%%%%%%%%%%%%%%%%%%%%%%%%%%%%%%%%%
%%%%%%%%%%%%%%%%%%%%%%%%%%%%%%%%%%%%%%%%%%%%%%%%%%%%%%%%%% 

\begin{center}

\renewcommand{\thefootnote}{\fnsymbol{footnote}}

  {\Large\bf
Failure of the collinear expansion in 
calculation of the induced gluon emission 
\\
\vspace{.7cm}
  }
\renewcommand{\thefootnote}{\arabic{footnote}}
\medskip
  {\large
  P.~Aurenche$^a$, B.G.~Zakharov$^{b}$ and H.~Zaraket$^{c}$ }
  \bigskip

{\it
$^{a}$Laboratoire d'Annecy-le-Vieux de Physique Th\'eorique LAPTH,\\
B.P. 110, F-74941 Annecy-le-Vieux Cedex, France\\
$^{b}$L.D. Landau Institute for Theoretical Physics,
        GSP-1, 117940,\\ Kosygina Str. 2, 117334 Moscow, Russia\\
$^{c}$Lebanese University Faculty of Sciences (I),\\
 Hadeth-Beirut, Lebanon
\vspace{1.7cm}}

  {\bf Abstract}
\end{center}
{
\baselineskip=9pt
We demonstrate that the collinear expansion fails in the case of 
gluon emission from a fast quark produced
in $eA$ DIS. In this approximation the $N\!=\!1$ rescattering contribution 
to the gluon spectrum vanishes. 
We show that
the higher-twist approach by Guo, Wang and Zhang \cite{W1,W2}
based on the collinear expansion is wrong.
The nonzero gluon spectrum obtained in \cite{W1,W2}
is a consequence of unjustified neglecting some important terms
in the collinear expansion.
\vspace{.5cm}
}

%-------------------------------------------------------------
\noindent{\bf 1}.
Since the early nineties considerable theoretical efforts 
have been made to study the induced gluon emission from fast partons
due to multiple scattering in cold nuclear matter and hot quark-gluon
plasma (QGP). 
There have been developed several approaches to this phenomenon.
The well known BDMPS \cite{BDMPS} and GLV \cite{GLV1} approaches
are based on the time-ordered diagram technique.  
The BDMPS formalism is valid for massless partons in the limit of strong 
Landau-Pomeranchuk-Migdal (LPM) \cite{LPM} suppression when the number
of rescatterings $N\gg 1$. 
The GLV approach accounts for small number of rescatterings ($N\le 3$) 
and applies only to thin plasmas.
The higher-twist method due to Guo, Wang and Zhang (GWZ)
\cite{W1,W2} is based on the Feynman diagram formalism and 
collinear expansion.
It includes only $N\!=\!1$ rescattering.
The formalism has originally been derived for the gluon emission
from a fast quark produced in $eA$ DIS. 
The light-cone path integral (LCPI) formalism \cite{Z1} 
(see also \cite{Z_YAF,BSZ,Wied1}) is free from the restrictions of the
approaches \cite{BDMPS,GLV1,W1,W2}. It accurately treats  
the mass effects, and applies at arbitrary 
LPM suppression. 

The LCPI \cite{Z1} and BDMPS \cite{BDMPS} approaches become 
equivalent at strong LPM suppression \cite{BSZ,Wied1}. The predictions
of the GLV \cite{GLV1} formalism can be obtained in the LCPI approach
\cite{Z1}
by expanding the gluon spectrum in the density of the medium.
However, the status of the GWZ approach \cite{W1,W2} is not clear.
The BDMPS-GLV-LCPI approaches in their original formulations 
neglect the quantum nonlocality in production of the fast partons.
In the case of gluon emission in the QGP produced in 
$AA$-collisions this
approximation is justified by the fact that the quantum nonlocality
is $\sim 1/E$ ($E$ is the energy of the fast parton), and can be safely
neglected at large $E$. 
In the GWZ approach to $eA$ DIS 
the nonlocal fast quark production and gluon emission have been treated 
on even footing.
However, as we will show below in the applicability region
of the GWZ formalism the quantum nonlocality in the quark production
in $eA$ DIS, similarly to
$AA$-collisions, is not important for gluon emission.
For this reason one could expect that the GWZ gluon spectrum 
should coincide with $N\!=\!1$ contribution in 
the LCPI formalism. But the two spectra are not identical.
Indeed, the GWZ gluon spectrum 
for the Gaussian nuclear density 
$n_{A}(r)\propto \exp{(-r^{2}/2R_{A}^{2})}$ at $z\ll 1$ (hereafter $z=\omega/E$, 
where $\omega$ is the gluon energy and $E$ is the 
struck quark energy) can be written as
(up to unimportant factors)
\beq
\frac{dP_{GWZ}}{dz}\propto\! \alpha_{s}^{2}n_{A}(0)R_{A}
P_{Gq}(z)
\int
\frac{dp^{2}}{p^{4}}  xG_{N}(x,p^{2}) 
\left\{
1-
\exp{
\left
[-\frac{p^{4}R_{A}^{2}}{4E^{2}z^{2}(1-z)^{2}}\right]}
\right\}\,,
\label{eq:10}
\eeq
where $P_{Gq}=C_{F}[1+(1-z)^{2}]/z$, $x\ll 1$, $G_{N}$ is the nucleon
gluon density. To leading order in $\alpha_{s}$ 
$xG_{N}(x,Q^{2})\approx\frac{3\alpha_{s}C_{F}}{\pi} 
\ln{(Q^{2}/\mu^{2})}$ ($Q^{2}\sim p^{2}$, $\mu$ is infrared
cutoff). 
On the other hand  
for the $N\!=\!1$ gluon spectrum in the LCPI approach one can obtain
for massless partons
\beq
\frac{dP_{LCPI}}{dz}\propto\! \alpha_{s}^{3}n_{A}(0)R_{A}
P_{Gq}(z)
\int
\frac{dp^{2}}{p^{2}(p^{2}+ \mu^{2})} \left\{
1-
\exp{\left[-\frac{p^{4}R_{A}^{2}}{4E^{2}z^{2}(1-z)^{2}}\right]}
\right\}\,.
\label{eq:20}
\eeq
One can see that the integrand in (\ref{eq:20}) does not have any 
logarithmic factor which could be interpreted as 
the nucleon gluon density entering (\ref{eq:10}).

In the present paper we clarify the situation
with the discrepancy between the predictions of the GWZ and LCPI approaches.
We demonstrate that the approximations used in \cite{W1,W2} 
really lead to a disagreement with the LCPI approach.
However, contrary to the results of \cite{W1,W2} the consistent use of 
the method
of \cite{W1,W2} gives a vanishing gluon spectrum.
This fact is a consequence of failure of the collinear expansion.
We show that the authors of Refs. \cite{W1,W2} obtained
the nonzero spectrum just due to unjustified neglecting some important
terms in the collinear expansion.

\vspace{.1cm}
\noindent{\bf 2}.
As in \cite{W1,W2},  we consider $eA$ DIS (we will discuss
the case when the virtual photon strikes out a quark with charge $e_{q}$).
As usual we choose the virtual
photon momentum in the negative $z$ direction, and describe 
the $0$ and $3$ components of the four-vectors 
in terms of the light-cone variables $y^{\pm}=(y^{0}\pm y^{3})/\sqrt{2}$.  
In the laboratory frame
the photon energy reads $\nu=Q^{2}/2m_{N}x_{B}$ ($x_{B}$ is 
the Bjorken variable).  The $p^{-}$ momentum of the struck quark
equals $\sqrt{2}\nu(1+x_{B}/2m_{N}\nu)$. The transverse momentum integrated 
distribution
for the $gq$ final state in $eA$ DIS can be described in terms of the  
semi-inclusive nuclear hadronic tensor 
$dW^{\mu\nu}_{A}/dz$.
To leading order in $\nu$ the spin effects in the final-state rescatterings of 
fast partons can be neglected. It ensures that the spin structure of 
$dW^{\mu\nu}_{A}/dz$ is the same as for the usual hadronic tensor 
$W^{\mu\nu}_{N}$
in $eN$ DIS. It allows one to write the semi-inclusive
nuclear hadronic tensor as 
$dW^{\mu\nu}_{A}/dz=-e^{2}_{q}g^{\mu\nu}_{T}df_{A}/dz$, where
$df_{A}/dz$
is the semi-inclusive quark distribution of the target nucleus. 
Neglecting the EMC and shadowing
effects it can be written as (we suppress the arguments $x_{B}$, $Q$ and $z$) 
\beq
\frac{df_{A}}{dz}=
\int d\rb n_{A}(\rb)
\frac{df_{N}(\rb)}{dz}\,,
\label{eq:30}
\eeq 
where $df_{N}(\rb)/dz$ is the in-medium semi-inclusive quark distribution
for a nucleon located at $\rb$, and $n_{A}(\rb)$ is the nucleus number density. 

Let us first discuss the evaluation of the in-medium 
semi-inclusive quark distribution in the LCPI formalism.
We will only concentrate on the aspects which are important
for comparison with the higher-twist method. Particularly
interesting is the question on the quantum nonlocality of the fast 
quark production in $eA$ DIS. 
Our consideration will be physical and diagrammatic.
A more formal analysis will be given in further
detailed publication.
In the LCPI approach \cite{Z1} the matrix element of the $q\rightarrow gq'$
in-medium transition is written in terms of the wave 
functions of the initial quark and final quark and gluon in the 
nucleus color field (we omit the color factors and indices)
\beq
\langle gq'|\hat{S}|q\rangle=ig\int\! dy 
\bar{\psi}_{q'}(y)\gamma^{\mu}A_{\mu}^{*}(y)\psi_{q}(y)\,.
\label{eq:40}
\eeq
Each quark wave function in (\ref{eq:40}) is written as 
$
\psi(y)=\exp(-ip^{-}y^{+})
\hat{u}_{\lambda}
\phi(y^{-},\vec{y}_{T})/\sqrt{2p^{-}}$,
where $\lambda$ is quark helicity,
$\hat{u}_{\lambda}$ is the Dirac spinor operator.
The $y^{-}$ dependence of the transverse wave functions
$\phi$ is governed by the two-dimensional 
Schr\"odinger equation 
\beq
i\frac{\partial\phi(y^{-},\vec{y}_{T})}{\partial
y^{-}}={\Big\{}\frac{
[(\vec{p}_{T}-g\vec{A}_{T})^{2}
+m^{2}_{q}]} {2\mu}
+gA^{+}{\Big\}}
\phi(y^{-},\vec{y}_{T})\,
\label{eq:60}
\eeq
with the Schr\"odinger ``mass'' $\mu=p^{-}$.
The wave function of the emitted
gluon can be represented in a similar way.

The $y^{-}$ evolution of the transverse wave functions can be written in terms
of the Green's function for the Schr\"odinger equation (\ref{eq:60})
\beq
\phi(y^{-}_{2},\vec{y}_{T,2})=
\int d\vec{y}_{T,1} {\cal{K}}(\vec{y}_{T,2},y^{-}_{2}|\vec{y}_{T,1},y^{-}_{1})
\phi(y^{-}_{1},\vec{y}_{T,1})\,.
\label{eq:70}
\eeq
It allows one to represent the matrix element in terms of the transverse 
Green's functions ${\cal{K}}$.
In the LCPI approach \cite{Z1} the path integral 
representation for the Green's functions is a crucial step 
in the final stage of calculations. It allows to perform averaging
over the medium potential at the level of the integrand in the 
double path integral representation of the gluon spectrum.
This method turns out to be very convenient in accounting for 
arbitrary number of rescatterings.
But for comparison with the higher-twist method we need only the $N\!=\!1$ 
rescattering contribution. It can be analyzed without using the path integral
representation of the Green's functions. 
For calculation of the $N\!=\!1$ term it is enough 
to have the perturbative expansion of the Green's functions 
to the second order in 
the external potential.
One can show that for gauges with potential 
vanishing at large
distances (say, covariant gauges, or Coulomb gauge) one can ignore the 
transverse component $\vec{A}_{T}$. Then, the second order expansion
of ${\cal{K}}$ can be written as
\bea
{\cal{K}}(\vec{y}_{T,2},y^{-}_{2}|\vec{y}_{T,1},y^{-}_{1})
=K(\vec{y}_{T,2},y^{-}_{2}|\vec{y}_{T,1},y^{-}_{1})-ig
\int dy^{-}
d\vec{y}_{T}
K(\vec{y}_{T,2},y^{-}_{2}|\vec{y}_{T},y^{-})A^{+}(\vec{y}_{T},y^{-})
\nonumber\\
\times 
 K(\vec{y}_{T},y^{-}|\vec{y}_{T,1},y^{-}_{1})
-g^{2}
\int\limits_{y^{-}>z^{-}} dy^{-}
d\vec{y}_{T}
\int dz^{-}
d\vec{z}_{T}
K(\vec{y}_{T,2},y^{-}_{2}|\vec{y}_{T},y^{-})A^{+}(\vec{y}_{T},y^{-})
\nonumber\\
\times K(\vec{y}_{T},y^{-}|\vec{z}_{T},z^{-})
A^{+}(\vec{z}_{T},z^{-})
K(\vec{z}_{T},z^{-}|\vec{y}_{T,1},y^{-}_{1})\,,
\label{eq:80}
\eea
where $K$ is the free Green's function when the external field is absent.
The expansion (\ref{eq:80}) allows one to represent
diagrammatically the $N\!=\!1$ 
induced gluon emission in $eA$ DIS by the set of diagrams like shown in 
Fig.~1 in which the horizontal solid line corresponds to $K$ ($\rightarrow$)
and $K^{*}$ ($\leftarrow$), the gluon line shows the correlator
$\langle A^{+}(y_{1})A^{+}(y_{2})\rangle$ in the nucleus,
and the vertical dashed line shows the transverse density
matrices of the final partons at very large $y^{-}$.
The graphs like Fig.~1a describe 
the real induced $q\rightarrow gq$ transition, and the  
virtual graphs like Fig.~1b, coming from the 
interference of the second and zeroth order terms on the right-hand side 
of (\ref{eq:80}), correspond to the unitarity correction due to interference of
of the induced and vacuum gluon emission.

The typical difference in the coordinate $y^{-}$ for the 
upper and lower $\gamma^{*}qq$ vertices 
in Fig.~1
(which gives the scale of the 
quantum nonlocality of the fast quark production)
is given by the well known Ioffe
length $L_{I}=1/m_{N}x_{B}$. For the usual
nucleon quark distribution $L_{I}$ is the dominating scale
in the Collins-Soper formula \cite{Soper} 
$f_{N}=\frac{1}{4\pi}\int dy^{-}e^{ix_{B}P^{+}y^{-}}
\langle N|\bar{\psi}(-y^{-}/2)\gamma^{+}\psi(y^{-}/2)|N\rangle$.
Contrary to 
the final state with single quark for the $gq$ final state the 
integration over the $y^{-}$ coordinate
of the $\gamma^{*}qq$ vertex is now affected by the integration over the 
positions of rescatterings and the $q\rightarrow gq$ splitting.
But for moderate $x_{B}$ when $L_{I}\ll R_{A}$ 
one can neglect the effect of rescatterings on the integration over $y^{-}$.
One can easily show that for production of the final $gq$ 
states with $M^{2}_{gq}\ll Q^{2}$ the restriction on $y^{-}$
from the splitting point can also be ignored.
Indeed, the typical scale in integrating over
the splitting points is given by the gluon formation length $L_{f}\sim
\nu/M^{2}_{gq}$ which is much bigger than $L_{I}$ at $M^{2}_{gq}\ll Q^{2}$
(this is true for both the vacuum DGLAP and the 
induced gluon emission).
Also, at 
$L_{I}\ll L_{f}$
one can take for the lower limit of the integration over the 
splitting points for the upper and lower parts of the diagrams in Fig.~1
the position of the struck nucleon. 
After this 
simplifications the quark production and gluon emission become mutually 
independent, and 
the 
${df_{N}(\rb)}/{dz}$
can be approximated by the factorized form 
\beq
\frac{df_{N}(\rb)}{dz}\approx f_{N}\frac{dP(\rb)}{dz}\,,
\label{eq:90}
\eeq 
where the quark distribution $f_{N}$ stems from the left parts of the diagrams,
and the gluon spectrum $dP/dz$ is described by the right parts of 
the diagrams evaluated neglecting
the quantum nonlocality of the fast quark production.

One can make some more simplifications.
If one neglects the multiquark configurations
in the nucleus both the vector potentials in the gluon correlators 
should belong to the same nucleon. For this reason the typical
separation of the arguments in the correlators  
is of the order of the nucleon radius, $R_{N}$. 
%%%%%%%%%%%%%%%%%%%%%%%%%%%%%%%%%%%%%%%%%%%%%%%%%%%
The restriction  $|y_{1}^{-}-y_{2}^{-}|\lsim R_{N}$ allows one
to replace the fast parton propagators between the gluon fields
in the graphs like Fig.~1b by $\delta$ 
functions in impact parameter space. This approximation is valid 
for parton energy $\gg 1/R_{N}$.
It follows from the Schr\"odinger diffusion
relation for the parton transverse motion $\rho^{2}\sim L/E$.
Also, the smallness of the fast parton diffusion radius at the 
longitudinal scale $\sim R_{N}$ allows one
to replace in other transverse Green's functions the $y^{-}$
coordinates by the mean values of the arguments of the vector potentials
in the gluon correlators. 
This approximation corresponds to a picture with 
rescatterings of fast partons on zero thickness scattering centers (nucleons).
The inequality $\omega \gg 1/R_{N}$ for the emitted
gluon is equivalent to $R_{N}\ll L_{f}$.
For this reason, in the picture of thin nucleons
the contribution of the graphs like Fig.~1c,d with gluon correlators 
connecting the initial quark and final quark or gluon 
can be neglected since they are suppressed by the small factor
$R_{N}/L_{f}$. 
These approximations have been used in the original formulation of
the LCPI approach \cite{Z1} (the BDMPS \cite {BDMPS} and GLV 
\cite{GLV1} approaches use them as well).
Note that (similarly to the case of QGP \cite{AZ}) each gluon correlator 
appears only in the form of an integral over $\Delta y^{-}=y^{-}_{2}-y^{-}_{1}$
and at $y^{+}_{2}=y^{+}_{1}$. One can easily show that 
this ensures  gauge invariance 
of the result (to leading order in $\alpha_{s}$).

The starting point of the higher-twist approach \cite{W1,W2} to the
gluon emission in $eA$ DIS is the representation of the nuclear 
hadronic tensor for the $\gamma^{*}A\rightarrow gqX$ transition in 
terms of the diagrams like shown 
in Fig.~2. The lower soft part is expressed in terms of the matrix element
$
\langle A|\bar{\psi}(0)A^{+}(y_{1})A^{+}(y_{2})\psi(y)|A\rangle
$, and  the upper hard parts are calculated perturbatively. 
In the calculation of the hard parts to leading order 
in the struck quark energy the integrations over the $y^{+}$
coordinates of the vertices give conservations of the large
$p^{-}$ momentum components of fast partons. 
At the same time the integration over 
the $p^{-}$ momentum components of the final partons ensures that
to leading order in the struck quark energy all the $y^{+}$ coordinates 
in the soft part can be set to zero.
Due to conservation of the large $p^{-}$ momenta of fast partons 
in the Feynman propagators only the Fourier components with 
$p^{-}>0$ are important. It means that the Feynman propagators are 
effectively reduced to the retarded (in $y^{-}$ coordinate) ones. 

One can easily demonstrate that the Feynman diagram treatment of \cite{W1,W2}
is equivalent to that in terms of the transverse Green's functions. 
Indeed, using 
the representation 
\beq
K(\vec{y}_{T,2},y^{-}_{2}|\vec{y}_{T,1},y^{-}_{1})=
i\int \frac{dp^{+}d\vec{p}_{T}}{(2\pi)^{3}}
\frac
{\exp{[
-ip^{+}(y^{-}_{2}-y{-}_{1})+i\vec{p}_{T}(\vec{y}_{T,2}-\vec{y}_{T,1})
]}}{p^{+}-
\frac{\vec{p}_{T}^{\,2}+m^{2}}{2p^{-}}+i0}
\label{eq:110}
\eeq
one can write the retarded quark propagator as
\bea
G_{r}(y_{2}-y_{1})=\frac{1}{4\pi}
\int_{0}^{\infty}\frac{dp^{-}}{p^{-}}
e^{-ip^{-}(y^{+}_{2}-y^{+}_{1})}\left[
\sum\limits_{\lambda}\hat{u}_{\lambda}\bar{\hat{u}}_{\lambda}
K(\vec{y}_{T,2},y^{-}_{2}|\vec{y}_{T,1},y^{-}_{1})
\right.
\nonumber\\
\left.
+i\gamma^{-}\delta(y^{-}_{2}-y^{-}_{1})
\delta(\vec{y}_{T,2}-\vec{y}_{T,1})\right]
\,.
\label{eq:120}
\eea
Here $\hat{u}_{\lambda}$ and $\bar{\hat{u}}_{\lambda}$ act on the variables
with indices 2 and 1, respectively. The last term in (\ref{eq:120}) 
is the so-called contact term. It does not propagate
in $y^{-}$ and can be omitted in calculating the 
nuclear final-state interaction effects for fast partons. One can easily 
verify that
using (\ref{eq:120}) and a similar representation for the gluon 
propagator
the hard parts in the higher-twist method can be represented in terms
of the transverse Green's functions as it is done in the LCPI treatment.
Thus, the hard parts in the approach 
\cite{W1,W2} should agree with that of the LCPI formalism.

\vspace{.1cm}
\noindent{\bf 3.}
The calculation of the diagrams like shown in Fig.~1 
is simplified by noting that the free transverse Green's function for a parton
with mass $m$ can be 
written as
\beq
K(\vec{y}_{T,2},y^{-}_{2}|\vec{y}_{T,1},y^{-}_{1})=
\theta(y^{-}_{2}-y^{-}_{1})\sum_{p_{T}}
\phi_{p_{T}}(\vec{y}_{T,2},y^{-}_{2})
\phi_{p_{T}}^{*}(\vec{y}_{T,1},y^{-}_{1})\,,
\label{eq:130}
\eeq
where $\phi_{p_{T}}(\vec{y}_{T},y^{-})\!=\!
\exp{[i\vec{p}_{T}\vec{y}_{T}-iy^{-}(\vec{p}_{T}^{\,2}+m^{2})/2p^{-}]}$
 is the plane wave solution to the
Schr\"odinger equation for $A^{\mu}=0$ with the transverse momentum 
$\vec{p}_{T}$. It allows one to represent the upper and lower parts of
the diagrams shown in Fig.~1 in the form
$
\int d y^{-}d\vec{y}_{T}
 \phi_{q'}^{*}(\vec{y}_{T},y^{-})
\phi_{g}^{*}(\vec{y}_{T},y^{-})
\phi_{q}(\vec{y}_{T},y^{-})
$
where the outgoing and incoming wave functions have the form of the plane
waves with sharp change of the transverse momentum at the points of 
interactions with the external gluon fields.
This method has previously been used in \cite{Z_kin} for investigation
of the kinematical effects. If one ignores the 
finite kinematical limits it reproduces the first order in the density 
term of the full LCPI expression for the induced gluon spectrum
obtained in \cite{Z_OA}.
We have checked that all the hard parts evaluated with the help of the 
plane waves agree with that obtained in Refs. \cite{W1,W2}. 

The sum of the complete set of the diagrams contributing to the $N\!=\!1$
spectrum can be written as \cite{Z_OA,Z_kin}
\beq
\frac{dP(\rb)}{dz}=\int\limits_{r_{3}}^{\infty} d\xi 
n_{A}(\vec{r}_{T},r_{3}+\xi)\frac{d\sigma(z,\xi)}{dz}\,.
\label{eq:140}
\eeq
Here $d\sigma(z,\xi)/dz$ is the cross section of gluon emission
from a fast quark produced at distance $\xi$ from the scattering nucleon.
At $z\ll 1$ (we consider the soft gluon emission
just to simplify the formulas) for massless partons it reads 
\cite{Z_OA,Z_kin} 
\beq
\frac{d\sigma (z,\xi)}{dz}=
\frac{2\alpha_{s}^{2} P_{Gq}(z)}{C_{F}}
\int 
d\vec{p}_{T}
\frac{d\vec{k}_{T}}{\vec{k}_{T}^{\,2}}  
\frac{xdG(k^{2}_{T},x)}{d\vec{k}_{T}}
{H(\vec{p}_{T},\vec{k}_{T},z,\xi) }\,,
\label{eq:150}
\eeq
\beq
H(\vec{p}_{T},\vec{k}_{T},z,\xi)\!=\!
\left[\frac{1}
{\vec{p}_{T}^{\,2}}-
\frac{(\vec{p}_{T}-\vec{k}_{T})\vec{p}_{T} }
{\vec{p}_{T}^{\,\,2}
(\vec{p}_{T}-\vec{k}_{T})^{2}
}\right]
\cdot
\left[1-
\cos\left(\frac{i\vec{p}_{T}^{\,2}\xi}{2Ez(1-z)}
\right)
\right]\,.
\label{eq:160}
\eeq
Here the limit $x\rightarrow 0$ is implicit, 
$dG(k^{2}_{T},x)/d\vec{k}_{T}$ is the unintegrated nucleon gluon density
which in leading order in $\alpha_{s}$ 
at $x\ll 1$ reads
\beq
\frac{dG(k^{2}_{T},x)}{d\vec{k}_{T}}=
\frac{1}{4\pi^{3}x}
\int 
d\vec{\rho} 
d\vec{y}_{T}
\exp{(-i\vec{k}_{T}\vec{\rho})} 
\langle \Psi_N| 
\nabla_{y_{T}}W^{a}(\vec{y}_{T}+\vec{\rho})
\nabla_{y_{T}}W^{a}(\vec{y}_{T})
|\Psi_N\rangle\,.
\label{eq:170}
\eeq
Here $W^{a}(\vec{y}_{T})=\int dy^{-} A^{+a}(y^{-},\vec{y}_{T})$
(the color index, $a$, is shown explicitly),
and $\Psi_N$ is the internal nucleon wave function normalized to unity.
One can show that the formula (\ref{eq:170}) being rewritten 
through the nucleon wave functions with relativistic 
normalization $\langle N'|N\rangle=2p^{+}(2\pi)^{3}\delta(p^{+'}-p^{+})
\delta(\vec{p}_{T}^{\,\,'}-\vec{p}_{T})$  
is reduced to the Collins-Soper definition
\cite{Soper} of the gluon density.
Eq. (\ref{eq:170}) can also be written as
\beq
\frac{dG(k^{2}_{T},x)}{d\vec{k}_{T}}=
\frac{N_{c}^{2}-1}{x32\pi^{4}\alpha_{s}C_{F}}
\int  d\vec{\rho}\exp{(-i\vec{k}_{T}\vec{\rho})}
{\nabla}^{2}\sigma(\rho)\,,
\label{eq:180}
\eeq
where $\sigma(\rho)$ is the well known dipole cross section
given by  
\beq
\sigma(\rho)=\frac{8\pi\alpha_{s}^{2}C_{F}}{N_{c}^{2}-1}
\int d\vec{y}_{T}
\langle \Psi_N| 
W^{a}(\vec{y}_{T})W^{a}(\vec{y}_{T})-
W^{a}(\vec{y}_{T}+\vec{\rho})
W^{a}(\vec{y}_{T})
|\Psi_N\rangle\,.
\label{eq:190}
\eeq
Using (\ref{eq:140})-(\ref{eq:170}) with the
phase $W^{a}$ calculated in the Born approximation one 
can obtain (\ref{eq:20}).

The collinear expansion corresponds to replacement of the hard part by
its second order expansion in $\vec{k}_{T}$ (we suppress all the arguments
except for $\vec{k}_{T}$)  
\beq
H(\vec{k}_{T})\approx
H(\vec{k}_{T}=0)+
\left.\frac{\partial H}
{\partial k^{\alpha}_{T}}\right|_{\vec{k}_{T}=0}k^{\alpha}_{T}
+
\left.\frac{\partial^{2} H}
{\partial k^{\alpha}_{T} \partial k^{\beta}_{T}}
\right|_{\vec{k}_{T}=0}\cdot
\frac{k^{\alpha}_{T}k^{\beta}_{T}}{2}
\,.
\label{eq:200}
\eeq
Then, to logarithmic accuracy 
$
d\sigma (z,\xi)/dz
\propto 
\int d\vec{p}_{T} 
x G(p^{2}_{T},x)
\nabla^{2}_{k_{T}}H(\vec{p}_{T},\vec{k}_{T},z,\xi)
|_{\vec{k}_{T}=0} 
$.
But from (\ref{eq:160}) one can easily obtain 
$\nabla_{k_{T}}^{2}H|_{\vec{k}_{T}=0}\!=0\!$.
It is also seen from averaging of the hard part over
the azimuthal angle of $\vec{k}_{T}$ which gives explicitly
\beq
\frac{1}{2\pi}\int d\phi_{k_{T}}
H(\vec{p}_{T},\vec{k}_{T},z,\xi)=
\frac{\theta(k_{T}-p_{T})}{\vec{p}_{T}^{\,2}}
\left[1-
\cos\left(\frac{i\vec{p}_{T}^{\,2}\xi}{2Ez(1-z)}
\right)
\right]\,.
\label{eq:220}
\eeq
Thus, contrary to the expected 
in the collinear approximation 
dominance of the region $k_{T}\lsim p_{T}$ only the 
region $k_{T}>p_{T}$ contributes to the gluon emission, and formal use of
the collinear expansion gives completely wrong result with zero gluon spectrum.

The fact that the $N\!=\!1$ gluon spectrum vanishes in the collinear
approximation agrees with prediction of the harmonic oscillator
approximation in the LCPI \cite{Z1} and BDMPS \cite{BDMPS} approaches. 
The full gluon spectrum in these approaches 
is expressed in terms of the Green's function of the Schr\"odinger equation
with an imaginary potential which is proportional to the dipole cross 
section. The corresponding Hamiltonian takes the harmonic oscillator form
for a quadratic parametrization $\sigma(\rho) = C\, \rho^2$.
The $N\!=\!1$ contribution to the gluon spectrum in the oscillator 
approximation should coincide
with prediction of the collinear expansion in the GWZ treatment \cite{W1,W2}
since the oscillator approximation in the LCPI and BDMPS approaches
is equivalent to the collinear approximation of \cite{W1,W2}.  
Indeed, the quadratic form of the dipole cross section 
corresponds to the vector potential 
approximated by the linear expansion 
$
A^{+}(y^{-},\vec{y}_{T}+\vec{\rho})\!\approx\!
A^{+}(y^{-},\vec{y}_{T})+
\vec{\rho}%\frac{\partial}{\partial \vec{y}_{T}}\,
\nabla_{y_{T}}
A^{+}(y^{-},\vec{y}_{T})\,
$
which can be traced back to the
collinear expansion in momentum space.
For a target of thickness $L$ in the oscillator approximation
the BDMPS and LCPI approaches for massless partons give the spectrum
\cite{BDMPS,BSZ,Z_OA}
\beq
\frac{dP_{OA}}{dz}=\frac{\alpha_{s}P_{Gq}(z)}{\pi}
\ln|\cos\Omega L|\,,
\label{eq:240}
\eeq
where
$\Omega=\sqrt{-i{C_{3} n}/{z(1-z)E}}$, $C_{3}=C C_{A}/C_{F}$.
Keeping only the first term in the expansion of the spectrum (\ref{eq:240}) in 
the density one gets \cite{Z_OA} 
\beq
\frac{dP_{OA}}{dz}\approx\frac{\alpha_{s}P_{Gq}(z)C_{3}^{2}n^{2}L^{4}}
{16\pi E^{2} z^{2}(1-z)^{2}}\,.
\label{eq:250}
\eeq
Since the right-hand side of (\ref{eq:250}) $\propto n^{2}$  
it corresponds to the $N\!=\!2$ rescatterings \cite{Z_OA}, and
the contribution of $N\!=\!1$ rescattering is absent. 
Note that from the point of view 
of the representation (\ref{eq:150}) the zero $N\!=\!1$ gluon spectrum in 
the oscillator 
approximation is a consequence of the fact that in this case 
$dG/d\vec{k}_{T}\propto \delta(\vec{k}_{T})$ (as one sees from (\ref{eq:180})).

\vspace{.1cm}
\noindent{\bf 4.}
Let us now discuss why the calculations of Refs. \cite{W1,W2} give nonzero
gluon spectrum. In \cite{W1,W2} the nonvanishing second derivative
of the hard part  comes from the graph shown in Fig.~2b (at $z\ll 1$).
The authors use for the integration variable in the hard part of
this graph the transverse momentum of the final gluon, $\vec{l}_{T}$.
The 
$\vec{l}_{T}$-integrated  hard part obtained in \cite{W2} (Eq. 15 of 
\cite{W2}) reads
(up to an unimportant factor)  
\beq
H(\vec{k}_{T})\propto \int \frac{d\vec{l}_{T}}
{(\vec{l}_{T}-\vec{k}_{T})^{2}} R(y^{-},y^{-}_{1},y^{-}_{2},
\vec{l}_{T},\vec{k}_{T})\,,
\label{eq:260}
\eeq
where
\bea
{R}(y^{-},y^{-}_{1},y^{-}_{2},
\vec{l}_{T},\vec{k}_{T})=\frac{1}{2}
\exp{
\left[i
\frac
{y^{-}(\vec{l}_{T}-\vec{k}_{T})^{2}-
(1-z)(y^{-}_{1}-y^{-}_{2})(\vec{k}_{T}^{\,2}
-2\vec{l}_{T}\vec{k}_{T})}{2q^{-}z(1-z)}
\right]
}\nonumber\\
\times\left[
1-\exp{
\left(i
\frac{
(y^{-}_{1}-y^{-})(\vec{l}_{T}-\vec{k}_{T})^{2}}
{2q^{-}z(1-z)}\right)}
\right]
\cdot
\left[
1-\exp{
\left(-i
\frac{
y^{-}_{2}(\vec{l}_{T}-\vec{k}_{T})^{2}}
{2q^{-}z(1-z)}\right)}
\right]
\label{eq:270}
\eea
is an analog of the last factor in the square brackets 
in (\ref{eq:160}) (our $z$ corresponds to $1-z$ in \cite{W1,W2}), 
the coordinates  
$y^{-}$, $y^{-}_{1,2}$ correspond to the quark interactions 
with the virtual photon and $t$-channel gluons. In calculating 
$\nabla_{k_{T}}^{2}H$
the authors differentiate only the singular factor
$1/(\vec{l}_{T}-\vec{k}_{T})^{2}$. But the omitted terms 
from differentiating the factor $R$
are important. After the $\vec{l}_{T}$ integration they almost completely 
cancel the contribution
from the $1/(\vec{l}_{T}-\vec{k}_{T})^{2}$ term. Indeed, after 
putting $y^{-}_{1}=y^{-}_{2}$ and changing the integration
variable 
$ \vec{l}_{T}\rightarrow (\vec{l}_{T}+\vec{k}_{T})$ the 
right-hand part of (\ref{eq:260}) does not depend on
$\vec{k}_{T}$ at all.
If one does not put $y^{-}_{1}=y^{-}_{2}$, there will be
some nonzero contribution to 
$\nabla_{k_{T}}^{2}H|_{\vec{k}_{T}=0}$
which,
however, is suppressed by the small factor $(R_{N}/L_{f})^{2}$. 
Keeping such contributions
does not make any sense since they  are clearly beyond predictive accuracy of
the approximations used in \cite{W1,W2}
\footnote{
Keeping the nonzero $y^{-}$ in the $\vec{l}_{T}$-,
$\vec{k}_{T}$-dependent terms, as it is done in \cite{W1,W2}, also
does not make sense  under the approximations of \cite{W1,W2}.}. 

\vspace{.1cm}
\noindent {\bf 5}. 
In summary, we have demonstrated that 
the collinear expansion
fails in the case of gluon emission from a fast massless quark produced
in $eA$ DIS. It gives a zero $N\!=\!1$ rescattering contribution 
to the gluon spectrum. This agrees with 
vanishing $N\!=\!1$ contribution to the gluon spectrum 
in the BDMPS \cite{BDMPS} and LCPI \cite{Z1} approaches
in the harmonic oscillator approximation which corresponds to the 
collinear expansion in momentum space.
The nonzero gluon spectrum obtained in \cite{W1,W2}
is a consequence of unjustified neglecting some important terms
in the collinear expansion.
The established facts demonstrate that the GWZ approach \cite{W1,W2}
is completely wrong. Its predictions 
for $eA$ DIS and jet quenching in $AA$ collisions 
do not make sense.

\vspace {.2 cm}
\noindent
{\large\bf Acknowledgements}

The work of BGZ  is supported 
in part by the grant RFBR
06-02-16078-a and the ENS-Landau program.

\newpage

%------------------------------------------------------------------
\begin{center}
{\Large \bf Figures}
\end{center}
%------------------------------------------------------------------

\begin{center}
\epsfig{file=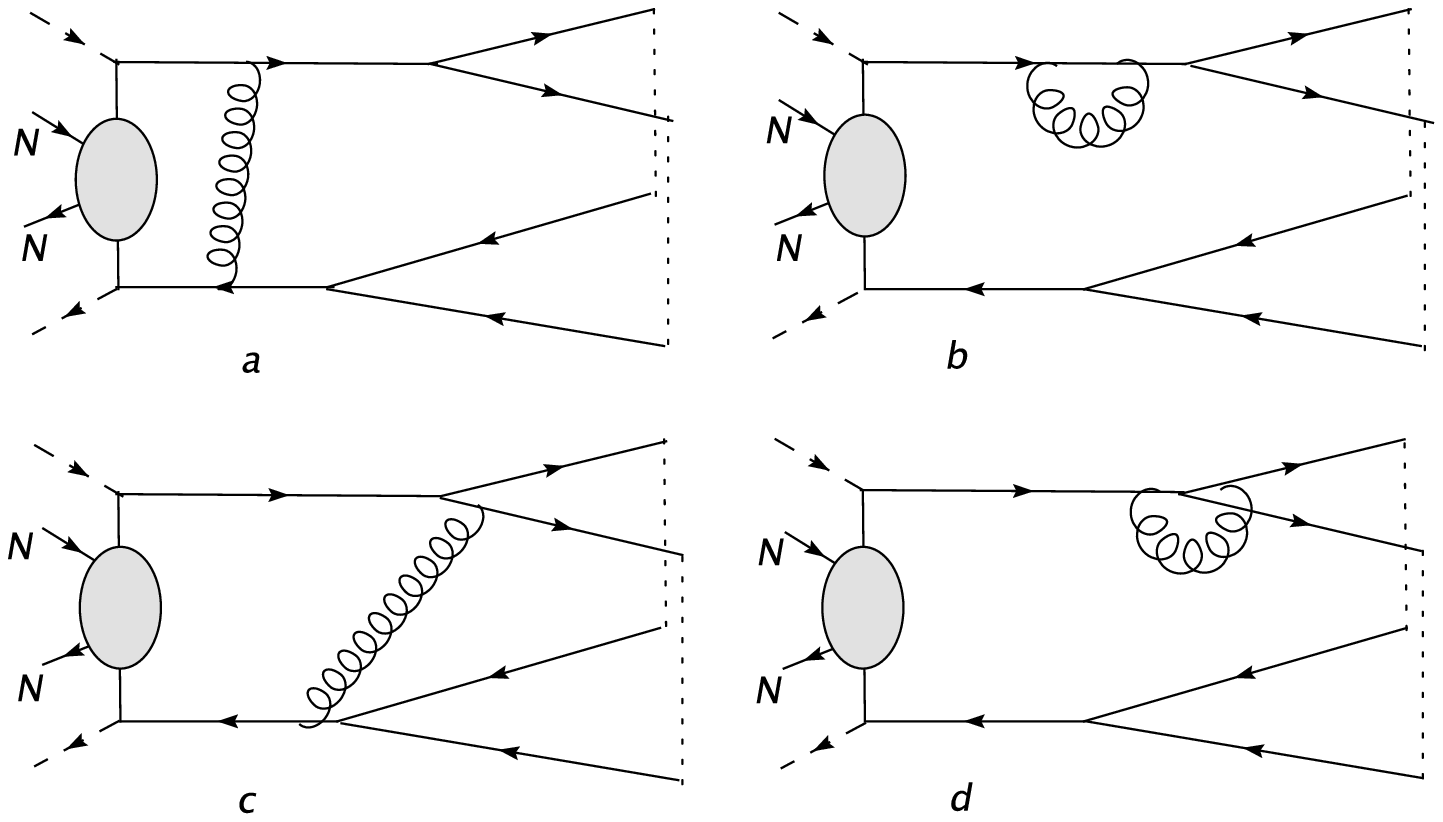,height=7cm,angle=0}
\end{center}
\begin{center}Figure 1. 
\end{center}

\begin{center}
\epsfig{file=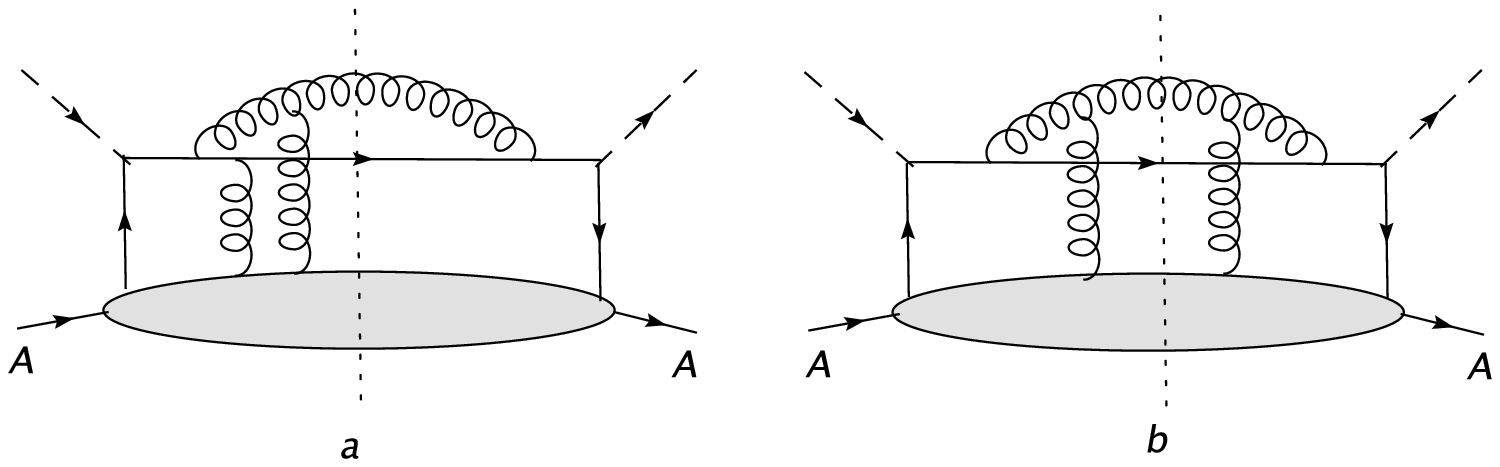,height=4.3cm,angle=0}
\end{center}
\begin{center}Figure 2. 
\end{center}

\end{document}